\documentclass[preprintnumbers,amsmath,amssymb,floatfix,9pt,one
column,prd,superscriptaddress,nofootinbib,showpacs,showkeys]{revtex4}
\usepackage{graphicx}
\usepackage{epsfig}
\usepackage{bm}
\usepackage{amsfonts}
\usepackage{amsfonts,amssymb,amsmath}
\usepackage[hang]{subfigure}

\def\ben{\begin{equation}}
\def\een{\end{equation}}

 \def\bd{\begin{document}} \def\ed{\end{document}}
\def\ds{\documentstyle} \let\fr=\frac \let\bl=\bigl \let\br=\bigr
\let\Br=\Bigr \let\Bl=\Bigl
\let\bm=\bibitem
\let\na=\nabla
\let\pa=\partial \let\ov=\overline
\newcommand{\be}{\begin{equation}}
\newcommand{\ee}{\end{equation}}
\def\ba{\begin{array}}
\def\ea{\end{array}}
\def\ft#1#2{{\textstyle{\frac{\scriptstyle #1}{\scriptstyle #2} } }}
\def\fft#1#2{{\frac{#1}{#2}}}
\def\del{\partial}
\def\vp{\varphi}
\def\sst#1{{\scriptscriptstyle #1}}
\def\oneone{\rlap 1\mkern4mu{\rm l}}
\def\td{\tilde}
\def\wtd{\widetilde}
\def\ie{{\it i.e.\ }}
\def\dalemb#1#2{{\vbox{\hrule height .#2pt
        \hbox{\vrule width.#2pt height#1pt \kern#1pt
                \vrule width.#2pt}
        \hrule height.#2pt}}}
\def\square{\mathord{\dalemb{6.8}{7}\hbox{\hskip1pt}}}
\newcommand{\ho}[1]{$\, ^{#1}$}
\newcommand{\hoch}[1]{$\, ^{#1}$}
\newcommand{\bea}{\setlength\arraycolsep{2pt} \begin{eqnarray}}
\newcommand{\eea}{\end{eqnarray}}
\newcommand{\ra}{\rightarrow}
\newcommand{\lra}{\longrightarrow}
\newcommand{\Lra}{\Leftrightarrow}
\newcommand{\bp}{\tilde \beta^\prime}
\newcommand{\tr}{{\rm tr} }
\newcommand{\Tr}{{\rm Tr} }
\def\0{{\sst{(0)}}}
\def\1{{\sst{(1)}}}
\def\2{{\sst{(2)}}}
\def\3{{\sst{(3)}}}
\def\4{{\sst{(4)}}}
\def\5{{\sst{(5)}}}
\def\6{{\sst{(6)}}}
\def\7{{\sst{(7)}}}
\def\8{{\sst{(8)}}}
\def\m{{\sst{(m)}}}
\def\n{{\sst{(n)}}}
\def\cA{{{\cal A}}}
\def\cB{{{\cal B}}}
\def\cF{{{\cal F}}}
\def\cG{{{\cal G}}}
\def\cH{{{\cal H}}}
\def\tV{\widetilde V}
\def\tW{\widetilde W}
\def\tH{\widetilde H}
\def\tE{\widetilde E}

\def\tF{\widetilde F}
\def\tA{\widetilde A}
\def\im{{{\rm i}}}
\def\tY{{{\wtd Y}}}
\def\ep{{\epsilon}}
\def\vep{{\varepsilon}}
\def\bD{{{\bar D}}}
\def\R{{{\mathbb R}}}
\def\C{{{\mathbb C}}}
\def\H{{{\mathbb H}}}
\def\CP{{{\mathbb C}{\mathbb P}}}
\def\RP{{{\mathbb R}{\mathbb P}}}
\def\Z{{{\mathbb Z}}}
\def\bA{{{\mathbb A}}}
\def\bB{{{\mathbb B}}}
\def\bC{{{\mathbb C}}}
\def\bD{{{\mathbb D}}}
\def\bE{{{\mathbb E}}}
\def\bZ{{{\mathbb Z}}}
\def\Re{{{\frak{Re}}}}
\def\Im{{{\frak{Im}}}}
\def\cosec{{\,\hbox{cosec}\,}}
\def\Gm{{\Gamma_{\!\! -}}}
\def\Gp{{\Gamma_{\!\! +}}}
\def\stan{{standard }}
\def\nonstan{{supernumerary }}
\def\p{{\partial}}
\def\kdel#1{{\fft{\del}{\del#1}}}

\def\bog{{Bogomolny }}
\def\om{{\omega}}

\newcommand{\nnr}{\nonumber \\}
\newcommand{\pd}{\partial}
\newcommand{\ud}{\textrm{d}}
\newcommand{\dTH}{T^{\prime \, 0}_\textrm{H}}
\newcommand{\dOi}{\Omega^{\prime \, 0}_i}
\newcommand{\bx}{{\bf x}}
\begin{document}

\title{p-wave holographic superconductors with Weyl corrections}
\author{\textbf{D. Momeni}}
\email{d.momeni@yahoo.com}
 \affiliation{Eurasian International Center for Theoretical Physics, Eurasian National University, Astana 010008, Kazakhstan}
\author{\textbf{ N. Majd}}
\email{naymajd@ut.ac.ir}
 \affiliation{Department of Engineering Science, Faculty of Engineering,
University of Tehran, Tehran, PO Box 11155-4563, Iran}

\author{\textbf{ R. Myrzakulov }}
\email{rmyrzakulov@gmail.com}
 \affiliation{Eurasian International Center for Theoretical Physics,  Eurasian National University, Astana 010008, Kazakhstan}

\begin{abstract}
We study the (3+1) dimensional  p-wave holographic  superconductors
with Weyl corrections both numerically and analytically. We describe
numerically the behavior of critical temperature $T_{c}$ with
respect to charge density $\rho$ in a limited range of Weyl coupling
parameter $\gamma$ and we find in general the condensation becomes
harder with the increase of
 parameter $\gamma$. In strong coupling limit of Yang-Mills theory, we show that the
minimum value of $T_{c}$ obtained from analytical approach is in
good agreement with the numerical results, and finally show how we
got remarkably a similar result in the critical exponent
$\frac{1}{2}$ of the chemical potential $\mu$ and the order
parameter$\langle J^1 _x \rangle$ with the numerical curves of
superconductors.

\end{abstract}
\pacs{ 11.25.Tq, 04.70.Bw, 74.20.-z}
 \keywords{Gauge/string duality; Classical black holes; Theories and models of superconducting states;
  Weyl corrections; P-wave holographic superconductors}

\newpage
\maketitle
\section{Introduction}
 Anti-de Sitter/ Conformal field theory (AdS/CFT) 
links a d- dimensional strongly coupled conformal field theory on
the boundary to a $(d+1)$- dimensional weakly coupled dual
gravitational description in the bulk\cite{maldacena, gubser,
hartnoll}.  It is necessary to couple  a complex scalar field with an Einstein-
Maxwell theory  to explain the simplest model for
holographic superconductors. In a holographic superconductor, below a critical temperature, the gauge symmetry
breaks and a black hole is constructed by the unstable developing
scalar hair near the horizon. According to the AdS/CFT
correspondence, the complex scalar field is dual to a charged
operator at the boundary, therefore a superconductor phase
transition will be occurred by both the $U(1)$ symmetry-breaking in
the gravity and a global $U(1)$ symmetry-breaking in the dual boundary
theory\cite{ weinberg}.\\
Holographic superconductors have been properly considered in
 two different models , one with
 an Abelian-Higgs model, which is the gravity dual of an $s$-wave
 superconductor with a scalar order parameter. This model has been
 studied by many authors see\cite{gubser, hartnoll,GR,herzog,albash1,albash2,horwitz,gubser2,keranen1,keranen2,chen,maeda}.
The other type uses a $SU(2)$ gauge theory \cite{pwave, roberts,ammon1,ammon2,ammon3,sonner,herzog2,pallab,gubser3}.\\
 Hartnoll et al.\cite{hartnoll} investigate further this
Abelian-Higgs model of superconductivity, and according to AdS/CFT
correspondence construct a $s$-wave holographic superconductor
solution with a scalar order parameter displaying the phase
transition process at the critical temperature $T_{c}$ below which
the charge condensate form. The Einstein-Yang-Mills (EYM) model of holographic
superconductors  constructed later by Gubser\cite{gubser4}, where
spontaneous symmetry-breaking solutions through a condensate of non
Abelian gauge fields is presented, and  also p-wave and $(p+ip)$-wave
backgrounds have been studied \cite{pwave}. In this case,the CFT have a global $SU(2)$ symmetry and hence three conserved
currents .
 The first effort on analytic methods in this topic
was the Herzog's work\cite{herzog3}, where critical exponent and the
expectation values of the dual operators was
attained.\\
Analytical studies of superconductors have been established in two
major methods: \textit{the small parameter perturbation theory}  as
in\cite{herzog3}, and \textit{the variational method}
\cite{analytic,me}. In the presented paper, we study the Weyl
corrected p-wave holographic superconductor composed of a
non-Abelian SU(2)  gauge field( the matter sector) and a black hole
background( the gravity sector) by using the variational method
giving only critical temperature $T_{c}^{Min}$. As it has been
mentioned in\cite{weyl}, for an Abelian gauge field and large range
of the Weyl coupling value $\frac{-1}{16} < \gamma < \frac{1}{24}$,
the universal relation for the critical temperature $T_{c} \approx
\sqrt[3]{\rho}$ has been found. In this paper we have explored the
same validity of the critical temperature relation in a non-Abelian
gauge field with Weyl
correction numerically and analytically.\\
The organization of this paper is as follows. In section II we
reconstruct the Weyl corrected superconductor's solution of the EYM
theory, which is dual to a p-wave superconductor. In section III we
present numerical results for condensation and critical temperature
of the holographic superconductor. Then we investigate the behavior
of critical temperature $T_{c}$ and dual of chemical potential $\mu$
with respect to Weyl coupling parameter $\gamma$, dual of charge
density $\rho$, and order parameter $\langle J_{x}^{1} \rangle$
analytically in section IV. The conclusion and some discussion are
given in section V.

\section{ Weyl corrected P-wave superconductors}
 In this section we study the holographic phase
transition for the probe $SU(2)$ Yang-Mills (YM) field
$A^{a}_{\mu}$ in a five dimensional space-time. The bulk action of
the Weyl gravity with an $SU(2)$ Yang-Mills in a five dimensional
spacetime is:
 \begin{eqnarray}
S=\int dt d^{4}x\sqrt{-g}\{\frac{1}{16\pi
G_{5}}(R+12)-\frac{1}{4g^2}F^{a}_{\mu\nu}F^{a\mu\nu}+\gamma
C^{\mu\nu\rho\sigma}F^{a}_{\mu\nu}F^{a}_{\rho\sigma}\}
\end{eqnarray}

Here $G_{5}$ is the five dimensional gravitational constant, $g$ is the Yang-Mills
coupling constant, and the negative cosmological constant is
satisfied by the factor $\frac{12}{l^2},\ \ l=1$ in the first parenthesis. The field strength component is given as below, where
$A_{\mu}^{a}$'s are the 
non-Abelian gauge fields.

 \begin{eqnarray}
 F^{a}_{\mu\nu}=\partial_{\mu}A^{a}_{\nu}-\partial_{\nu}A^{a}_{\mu}
 +\varepsilon^{abc}A^{b}_{\mu}A^{c}_{\nu},\{a=1,2,3,\mu=0,1,2,3\}
 \end{eqnarray}
 The Weyl's coupling $\gamma$
is limited such that its value is in the interval
$-\frac{1}{16}<\gamma<\frac{1}{24}$, (for more precise details see \cite{ritz}). In the probe limit by  neglecting  the
back reactions , the gravity sector is effectively
decoupled from the matter field's sector. In this probe limit, the background
metric is given by an AdS-Schwarzschild black hole:
\begin{eqnarray}
 ds^2=r^2(-fdt^2+dx^idx_i)+\frac{dr^2}{r^2f}
\end{eqnarray}
where:
\begin{eqnarray}
f=1-(\frac{r_{+}}{r})^4
\end{eqnarray}
The black hole horizon is $r=r_{+}$. The Smarr-Bekenstein-Hawking
temperature of the black hole is determined by the Schwarzschild
radius as $T=\frac{r_{+}}{\pi }$. This is the same as the temperature of the
conformal field theory on the boundary of the AdS spacetime.
Applying the Euler-Lagrange equation, we can derive the generalized
Yang-Mills equation as\cite{weyl}:
\begin{eqnarray}\label{GYME1}
\nabla_{\mu}\left( F^{a\mu\nu} - 4\gamma C^{\mu\nu\rho\sigma}
F^{a}_{\rho\sigma} \right) =-\epsilon^{a}_{bc}A^{b}_{\mu}F^{c\mu\nu}
+ 4\gamma
C^{\mu\nu\rho\sigma}\epsilon^{a}_{bc}A^{b}_{\mu}F^{c}_{\rho\sigma}
\end{eqnarray}
where $C_{\mu\nu\rho\sigma}$ is the Weyl tensor and has the
following nonzero components in $AdS^{5}$:
\begin{eqnarray}
\label{WeylT} C_{0i0j}=f(r)r_{+}^{4} \delta_{ij},~~
C_{0r0r}=-\frac{3 r_{+}^{4}}{ r^{4}},~~ C_{irjr}=-\frac{r_{+}^{4}}{
r^{4} f(r)} \delta_{ij},~~ C_{ijkl}=r_{+}^{4} \delta_{ik}
\delta_{jl}.
\end{eqnarray}

For realization of a holographic p-wave superconductor we take the following anstaz for Yang-Mills gauge field
\cite{gauge}:
 \begin{eqnarray}
 A=\varphi(r)\sigma^{3}dt+\psi(r)\sigma^1 dx
 \end{eqnarray}
 ($\sigma^{i}$ Pauli's matrixes). The condensation
of $\psi(r)$ breaks the $SU(2)$ symmetry and the final state is the
superconductor phase transition. The gauge function $\psi(r)$ is dual to
the $J^1 _x$ operator on the boundary, choosing $x$ axis as a special
direction, the condensation phase of $\psi(r)$ breaks the 
symmetry and leads to a phase transition, which can be interpreted
as a $p$-wave superconductor phase transition on the boundary. The
resulting Yang-Mills equations for metric (3) are given by:
\begin{eqnarray}
\label{EOMr1} \left( 1 - \frac{24 \gamma r_{+}^{4}}{r^{4}}
\right)\varphi'' + \left( \frac{3}{r} + \frac{24 \gamma
r_{+}^{4}}{r^{5}} \right)\varphi'
- \left( 1 + \frac{8 \gamma r_{+}^{4}}{r^{4}} \right)\frac{\psi^{2}\varphi}{r^{4}f}=0\\
\label{EOMr2} \left( 1 - \frac{8 \gamma r_{+}^{4}}{r^{4}}
\right)\psi'' + \left[ \frac{3}{r} + \frac{f'}{f} - \frac{8 \gamma
r_{+}^{4}}{r^{4}} \left( -\frac{1}{r} + \frac{f'}{f} \right) \right]
\psi' + \left( 1 + \frac{8 \gamma r_{+}^{4}}{r^{4}} \right)
\frac{\varphi^{2}\psi}{r^{4}f^{2}}=0
\end{eqnarray}
where the prime denotes derivative with respect to $r$. It is
more conveint to work in terms of the dimensionless parameter
$z=\frac{r_{+}}{r}$, in which at the horizon $z=1$, and the boundary
at the infinity locates at $z=0$. Then the the equations of motion
(\ref{EOMr1}) and (\ref{EOMr2}) can be reexpressed as:
\begin{eqnarray}
\label{EOMz1} \left( 1 - 24 \gamma z^{4} \right)\varphi'' -
\frac{1}{z} \left( 1 + 72 \gamma z^{4} \right)\varphi'
- \left( 1 + 8 \gamma z^{4} \right)\frac{\psi^{2}\varphi}{f}=0\\
\label{EOMz2} \left( 1 - 8 \gamma z^{4} \right)\psi'' + \left[
-\frac{1}{z} + \frac{f'}{f} - 8 \gamma z^{4} \left( \frac{3}{z} +
\frac{f'}{f} \right) \right] \psi' + \left( 1 + 8 \gamma z^{4}
\right) \frac{\varphi^{2}\psi}{f^{2}}=0
\end{eqnarray}
where the prime now denotes derivative with respect to $z$. The
boundary conditions at infinity, i.e. $z\rightarrow 0 $, are:
\begin{eqnarray}
 \varphi\simeq \mu-\rho z^2\\
 \psi\simeq\psi^{(0)}+\psi^{(2)}z^2
 \end{eqnarray}
$\mu$ and $\rho$ are dual to the chemical potential and charge
density of the CFT boundary, $\psi^{(0)}$ and $\psi^{(2)}$ are dual
to the source term and expectation value of the boundary operator $J^1
_x$ respectively.
 Further to have a normalizable solution,
  we always set the source $\psi^{(0)}$ to zero.

\begin{figure}
\center{
\includegraphics[scale=1.2]{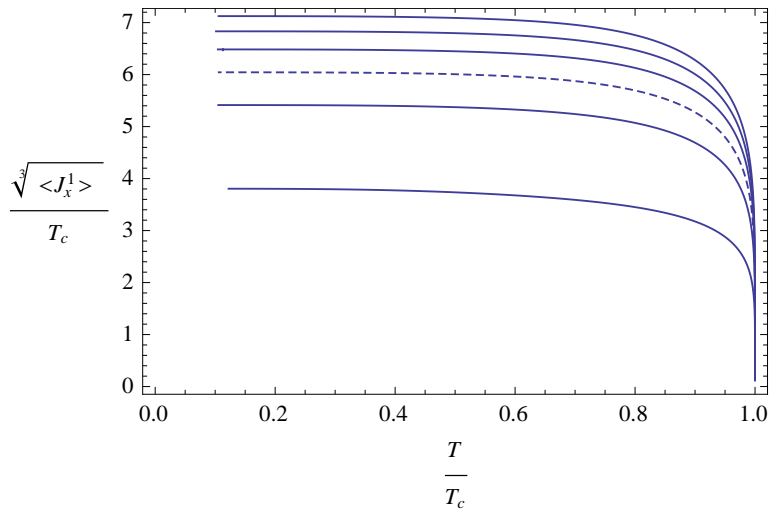}\hspace{1cm}
\caption{\label{Condense1} The condensation as a function of
temperature for the operators $<J_{x}^{1}>$. $\gamma=-0.06, -0.04,
-0.02, 0, 0.02, 0.04$ from top to bottom and the dotted line is just
the case $\gamma=0$.}}
\end{figure}

\begin{widetext}
\begin{table}[ht]
\begin{center}

\begin{tabular}{|c|c|c|c|c|c|c|}
         \hline
$~~\gamma~~$
&~~$-0.06$~~&~~$-0.04$~~&~~$-0.02$~~&~~$0$~~&~~$0.02$~~&~~$0.04$~~
          \\
        \hline
~~$T_{c}$~~ & ~~$0.1701\rho^{1/3}$~~ & ~~$0.1774\rho^{1/3}$~~ &
~~$0.1869\rho^{1/3}$~~& ~~$0.2005\rho^{1/3}$~~&
~~$0.2239\rho^{1/3}$~~& ~~$0.3185\rho^{1/3}$~~
          \\
        \hline
\end{tabular}
\caption{\label{Tc} The critical temperature $T_{c}$ for different
values of Weyl coupling parameter $\gamma$(Numerical results).}
\end{center}
\end{table}
\end{widetext}
\section{numerical treatment}
In this section we
will present numerical results for the condensation and critical
temperature due to the shooting method. From EOM's (\ref{EOMz1}) and
(\ref{EOMz2}) and the asymptotic behavior of $\psi$ and $\varphi$ at
infinity (12) and (13), we can obtain the regularity condition as:
\begin{eqnarray}
\label{reg}
 \varphi(1)=\varphi'(0)=\psi'(0)=\psi'(1)=0
\end{eqnarray}
combining boundary condition (12) and (13) with regularity condition
(\ref{reg}), we can solve EOM's (\ref{EOMz1}) and (\ref{EOMz2})
numerically by using a shooting method, and plot the FIG.1 to
demonstrate the condensation as a function of temperature for the
operator$< J^1 _x >$. The curve in FIG.1 is qualitatively similar to
that obtained in the holographic superconductors\cite{weyl,
horwitz2}, where the condensation of $< J^1 _x >$ goes to a constant
at zero temperature. As we can see from the FIG.1, it is easy to
find that the critical temperatures of Weyl corrected
superconductors is increasing as the parameter $\gamma$ varies in
the range of $-0.06$ to $0.04$, therefore we conclude that when $\gamma
< 0 $ the critical temperature is smaller and the formation of
scalar hair is harder and vice versa when $\gamma > 0
$.\\
We have also presented the critical temperature $T_{c}$ with
different values of the parameter $\gamma $ in the TABLE I.
According to the results in the TABLE I, we can conclude that by
minimizing the coupling parameter $\gamma $, the critical
temperature decrease smoothly .

\section{Analytical treatment}
In this section we compute the critical temperature and critical
exponent via an analytical method, which has been proposed recently
\cite{analytic}. In this method by defining appropriate equation
matching with field's boundary conditions, the field's EOM will be
transformed to Sturm-Liouville self adjoint form. Therefore according to the
general variational method to solve the Sturm-Liouville problem
(\cite{hartman} or appendix of \cite{analytic} ), the eigenvalue
$\lambda^{2}$ minimizing the Sturm-Liouwille equation can be found.
Using this minimum value of $\lambda$, one can obtain the minimum
critical temperature $T_{c}^{Min}$. In this section we calculate
this $T_{c}^{Min}$ and discuss the critical exponent.
\subsection{Critical temperature $T_C^{Min}$}

 Considering the non linear system (10,11). If there
is a second order continuous phase transition at the critical
temperature, the solution of the EOMs at the $T_C$ should be:
\begin{eqnarray}
\psi(z)=0,\varphi(z)=\lambda h_c(1-z^2)
 \end{eqnarray}
Here $\lambda=\frac{\rho}{h_c^3}$, $h_c$ is the radius of the
horizon corresponding to $T=T_c$. At a temperature slightly below
$T_c$, the EOM for $\psi$ becomes:
\begin{eqnarray}
 z^2\frac{d}{dz}((\frac{(1-z^4)}{2g^2z}+4\gamma h_c^3
z^3)\frac{d\psi}{dz})+\lambda^2[h_c^3(\frac{z}{2g^2}+4\gamma
z^5)\frac{1-z^4}{1+z^4}]\psi=0
\end{eqnarray}
It is appropriate to define:
\begin{eqnarray}
\psi(z)=\frac{\langle J^1 _x \rangle}{h}z^2 F(z)
\end{eqnarray}
Matching the boundary condition at the boundary $z=0$, we normalize
the function as $F(0)=1, F'(0)=0$. The equation for F(z) is:

\begin{eqnarray}
\frac{d}{dz}(k(z)\frac{dF(z)}{dz})-p(z)F(z)+\lambda^2q(z)F(z)=0
\end{eqnarray}

where:

\begin{eqnarray}
k(z)=\frac{z^3(1-z^4)}{2g^2 }+4\gamma h_c^3 z^7\\
p(z)=-2z^5(-\frac{2}{g^2}+16\gamma h_c^3)\\
 q(z)=h_c^3z^2(\frac{h_cz}{2g^2}+4\gamma z^5)\frac{1-z^4}{1+z^4}
\end{eqnarray}

The eigenvalue $\lambda$ minimizes the expression (18) is obtained
from the following functional:
\begin{eqnarray}
\lambda^2=\frac{\int_{0}^{1}(k(z)F'(z)^2+p(z)F(z)^2)dz}{\int_{0}^{1}q(z)F(z)^2dz}
\end{eqnarray}
To estimate it, we use the trial function $F(z)=1-\alpha z^2$. We
then obtain:
\begin{eqnarray}
\lambda_{\alpha}^2=\frac{2 g^2 \left(-1.6 h_c^3 \alpha ^2 \gamma +8.
h_c^3 \alpha  \gamma
   -5.33333 h_c^3 \gamma +\frac{0.533333 \alpha ^2}{g ^2}-\frac{
   \alpha }{g ^2}+\frac{0.666667}{g ^2}\right)}{h_c^3 \left(\alpha ^2
   \left(-0.21929 \gamma  g ^2-0.0151862\right)+\alpha  \left(0.628086
   \gamma  g ^2+0.052961\right)-0.485957 \gamma  g
   ^2-0.0568528\right)}
\end{eqnarray}
Which attains a minimum at $\alpha=0.304936$, and from the
$\lambda=\frac{\rho}{h_c^3}$ and $T_c=\frac{h_c}{\pi}$ the minimum
value of the critical temperature can be read as:
\begin{eqnarray}
T_c^{Min(\pm)}=0.256926 \sqrt[3]{\frac{-0.128539\pm0.3125 g ^2
\sqrt{\frac{1.90164
   \gamma  g ^2 \rho ^2 \left(1.00743 \gamma  g
   ^2+0.134769\right)+0.169187}{g ^4}}}{\gamma  g ^2}}
\end{eqnarray}

We know that in strong-coupling regime of the YM theory, the
quantities can be expanded in series of $\frac{1}{g}$ \cite{1/g}.
Since for some values of $g$, we may be have  $T_{c}^{Min(-)}<0$,
therefore only the $T_{c}^{Min(+)}$ is acceptable and can be read in
strong limit of order $\frac{1}{g^2}$ as:
\begin{equation}
    T_{c}^{Min(+)}\approx 0.1943040830 \rho^{\frac{1}{3}}+\frac{0.1497404
    (-0.128539\gamma+ 0.028931219\gamma\rho)}{\gamma^{2}\rho^\frac{2}{3}g^{2
    }}
\end{equation}
In prob limit by neglecting the back reaction, the large values of
the YM coupling is accessible. Comparing equation (25) with the
TABLE I we observe that the analytic value of the leading order
$T_{c}^{Min}\approx 0.1943040830 \rho^{\frac{1}{3}}$ obeys the well
known role $T_{c}\propto \rho^{\frac{1}{3}}$ and it is the lower
bound for tabulated values of $T_{c}$, given in TABLE I. According
to the numerical results of TABLE I, the minimum value of $T_{c}$
reads as:
\begin{equation}
T_{c(numeric)}^{Min(+)}\approx 0.1701 \rho^{\frac{1}{3}}
\end{equation}
which shows that the analytic values in relation (25) and the
numerical estimate in equation (26) are in good agreement with each
other. It seems that there is a deep relation between the strong
limit of YM part of the action and the analytical results of the
p-wave superconductors .

\subsection{Relation of $<J_x^1> -(\mu-\mu_c)$}
Now we want to know the behavior of the order parameter at $T_c$, by solving the equation for the scalar potential close to $T_c$,
therefore by substituting (13) in (10) we have:

\begin{eqnarray}
\frac{d}{dz}((-\frac{1}{2z g^2}+12\gamma
z^3)\frac{d\varphi}{dz})+(\frac{z}{h_c})^2(\frac{ z}{2(1-z^4)g^2
}+\frac{4\gamma z^5}{1-z^4 })(\frac{<J^1 _x>}{h_c}
)^2F(z)^2\varphi=0
\end{eqnarray}

Since near the critical point, the order parameter $<J^1 _x>$ is small, we can expand
$\varphi$as a series form of the small parameter as below:
\begin{eqnarray}
\varphi=\mu_c+<J^1 _x>\chi(z)+ ...
\end{eqnarray}
The boundary condition  imposes that
$\chi(z)$ to be $\chi(1)= 0$ .The EOM for $\chi(z)$ can be obtained
as below:
\begin{eqnarray}
\frac{d}{dz}((-\frac{1}{2z g^2}+12\gamma
z^3)\frac{d\chi(z)}{dz})+(\frac{z}{h_c})^2(\frac{ z}{2(1-z^4)g^2
}+\frac{4\gamma z^5}{1-z^4 })\frac{J^1 _x}{h_c^2}F(z)^2\mu_c=0
\end{eqnarray}
By integration from both sides of the (29), the EOM for $\chi(z)$
can be reduced to:
\begin{eqnarray}
(-\frac{1}{2z\zeta^2}+12\gamma
z^3)\frac{d\chi(z)}{dz}=-\mu_c\frac{<J^1 _x>}{h_c^4}\int z^2(\frac{
z}{2(1-z^4)g^2 }+\frac{4\gamma z^5}{1-z^4 })F(z)^2 dz
\end{eqnarray}
By regularity condition, we have $\chi'(0)=0$ ,
so we  take $F(z)=(1-z^2)(1-\alpha z^2)$. Now
$\varphi(z)$ can be expanded as:
\begin{eqnarray}
\varphi(z)\sim\mu-\rho z^2\approx\mu_c+<J^1
_x>(\chi(0)+\chi'(0)z+...)
\end{eqnarray}
 Now  from  (31), by comparing the
coefficients of $z^0$ term in both sides of the above formula, we
obtain:

\begin{equation}
\mu-\mu_c\approx\frac{(<J^1 _x>)^2 \mu_c}{34560 h^4 \gamma ^2 g
^4}(3003.22 \gamma ^{3/2} g ^3 \left(8 \gamma  g ^2+1\right)
\text{Li}_{2}\left(\frac{12. \sqrt{\gamma } g }{12.
   \sqrt{\gamma } g -2.44949}\right)+(9047.79+6359.31 ) \gamma ^2 g ^4)
\end{equation}
Where $\alpha=0.304936$ is a parameter minimizing the equation (23),
and $\text{Li}_{2}(z)$ gives the polylogarithm function.\\ This
critical exponent $\frac{1}{2}$ for the condensation value and
$(\mu-\mu_c)$ qualitatively match the numerical curves for
superconductors with Weyl corrections\cite{weyl}.

\section{conclusions}

In this letter we have investigated the Weyl corrected p-wave Holographic superconductors at the
probe limit using numerical and analytical
solutions. We obtained  the behavior of the critical temperature $T_{c}$ as a function of the dual charge density $\rho$ in different values of
Weyl coupling parameter $\frac{-1}{16}< \gamma
<\frac{1}{24}$. We have found that the critical temperature increases by growing the Weyl coupling parameter, therefore the
condensation becomes harder when $\gamma < 0 $, and vice versa when
$\gamma > 0$. As a final point, obtaining the critical exponent
$\frac{1}{2}$ for the chemical potential $\mu$ and order
parameter$\langle J^1 _x \rangle$ show a good agreement with the
numerical curves of Weyl corrected s-wave holographic superconductors. Furthermore, we
have shown that in the strong limit of YM theory, the analytical and numerical values of the minimum value
of the critical temperature are in good agreement.

\section{acknowledgement}
The authors would like to thank Jian-Pin Wu,
from Beijing Normal University (China), for helpful
suggestion and recommending useful references, and also we truthfully claim
that without his valuable numerical codes and results
the substantial improvements of this presentation and
outcomes would not have been achieved. Besides, we
thank the referee for good observations and kind guidelines.
NM would like to acknowledge the financial support
of University of Tehran for this research under grant No.
02/1/28450.

\end{document}